\documentclass[aps,prd,tightenlines,twocolumn,floatfix,nofootinbib,showpacs]{revtex4}
\usepackage{dcolumn}
\usepackage{graphicx}

\def\bml{\begin{mathletters}}
\def\eml{\end{mathletters}}
\def\ba{\begin{array}}
\def\ea{\end{array}}

\def\to{\rightarrow}

\newlength{\bredde}
\def\slash#1{\settowidth{\bredde}{$#1$}\ifmmode\,\raisebox{.15ex}{/}
\hspace*{-\bredde} #1\else$\,\raisebox{.15ex}{/}\hspace*{-\bredde} #1$\fi}

\begin{document}
\title{
Divergent chiral condensate in the quenched Schwinger model
}
\author{Poul H. Damgaard}
\affiliation{
The Niels Bohr Institute, Blegdamsvej 17, DK-2100 Copenhagen \O,
Denmark 
}
\author{Urs M. Heller}
\affiliation{
American Physical Society, One Research Road, Box 9000, Ridge,
NY 11961-9000, USA
}
\author{Rajamani Narayanan}
\affiliation{Department of Physics, Florida International University,
University Park, Miami FL 33199, USA
}
\author{Benjamin Svetitsky}
\affiliation{School of Physics and Astronomy, Raymond and Beverly Sackler
Faculty of Exact Sciences, Tel Aviv University, 69978 Tel Aviv, Israel
}

\begin{abstract} 
We calculate numerically the eigenvalue distribution of
the overlap Dirac operator in the quenched Schwinger
model on a lattice. The distribution does not fit any of the
three universality classes of spontaneous chiral symmetry breaking,
and its strong volume dependence indicates that the chiral condensate
in the quenched theory is an ill-defined and divergent quantity. When 
we reweight 
configurations with the Dirac determinant to study the theory with $N_f=1$, 
we obtain a
distribution of eigenvalues that is well-behaved and consistent with the 
theory
of explicit symmetry breaking due to the anomaly.
\end{abstract}
\maketitle


\section{Introduction}

Quantum electrodynamics in (1+1)-dimensions, the Schwinger model
\cite{Schwinger},
continues to play an important role as testing ground for field
theory ideas. In this paper we use the Schwinger model to study the quenched 
approximation
for the chiral condensate. The quenched chiral
condensate has long been believed, from various indirect arguments,
to be an ill-defined quantity in gauge theories in any number
of dimensions. Calculations in the Schwinger model are much easier than 
in higher dimensional gauge theories, and this is our 
motivation for the present study. One
must keep in mind, however, the fact that
spontaneous chiral symmetry breaking is prohibited in two dimensions by
the Coleman-Mermin-Wagner theorem \cite{MW}. For this reason we also
consider the unquenched theory, where chiral symmetry
is broken explicitly by the anomaly so that the chiral condensate
should be well defined. 

The quenched approximation, in which the fermion determinant is discarded when
generating the gauge field configurations, was first discussed
analytically in this theory by van den Doel \cite{Doel}.
Some of the subtleties were subsequently discussed in Refs.~\onlinecite{CK} 
and~\onlinecite{G}.
As we shall review in the next section, the
indications of disease in the quenched Schwinger model very much resemble 
those in
higher-dimensional
quenched theories, analyzed by means of quenched chiral perturbation
theory \cite{BG,DS}. This may not be surprising, since the bosonized
form of the Schwinger model \cite{bosonization} is a two-dimensional 
analogue
of a chiral Lagrangian. The trouble with the quenched chiral
condensate then stems, in both contexts, from
the famous double pole in the singlet correlation
function \cite{BG,S}. When analyzed in the finite-volume 
$\epsilon$-regime \cite{Smilga,LS}, the quenched chiral condensate
is seen to be plagued with a ``quenched finite-volume logarithm'' at one-loop
order in four dimensions \cite{D}.
Taken at face value, that is,  if one were to push the expansion beyond 
its region of 
validity, this could indicate a divergent
condensate. The
analysis of the quenched chiral condensate in the $\epsilon$-regime
has guided us in our present finite-volume calculations.

There have been many Monte Carlo calculations of the
chiral condensate in the Schwinger model, both quenched and unquenched (see 
\cite{NNV,Bietenholz1,Lang,Chandra,Bietenholz2,KN,GHR,Durr}
for recent work). The first
to study the quenched condensate via the distribution of the lowest
Dirac operator eigenvalues were Farchioni {\em et~al.}~\cite{Lang}. They
found quite an odd result---agreement with
one universality class of spontaneous chiral symmetry breaking at
small volumes, and with another at larger volumes. We shall return
to this issue in detail below. Kiskis and Narayanan \cite{KN} reconsidered 
the problem recently and found evidence for
a {\em divergent\/} quenched chiral condensate, with Dirac eigenvalues
that did not appear to fit any of the three possible chiral
symmetry breaking classes. We shall see how these last two papers can be 
reconciled.
In the process we shall consider volumes far exceeding what has been 
studied earlier. 

Our paper is organized as follows.
In the next section we briefly review the issues surrounding the quenched
chiral condensate, particularly in this two-dimensional setting, and
show how analytical arguments favor an ill-defined, divergent
quantity. In Sec.~III we turn to our Monte Carlo
simulations of the quenched theory.
We compare numerical results for the distribution
of Dirac operator eigenvalues with distributions based on the
three possible classes of spontaneous chiral symmetry breaking.
We show how the distributions drift as the volume is changed,
and how this explains the results of Ref.~\onlinecite{Lang}. Moreover,
we find that the distributions appear not to
converge to any fixed limit, indicating that the spectral
density $\rho(\lambda)$ does not attain a finite value at
$\lambda = 0$. By way of contrast, we present in Sec.~IV
numerical results for the unquenched theory ($N_f=1$), obtained by reweighting
the path integral with the Dirac determinant.
Our results there
are consistent with the detailed predictions of the $\epsilon$-regime
based on the explicit breaking of chiral symmetry due to the
anomaly \cite{LS}. Section V contains a brief summary of our
results.

\section{The quenched Schwinger model}

Analytically, one can treat quenching by means of the replica method
that considers $N_f$ identical fermion copies, and then sends
$N_f \to 0$ at the end. This can be done trivially in perturbation
theory at the fundamental level; in more than two dimensions
it can be done 
in chiral perturbation theory \cite{DS}. Because
the fermion determinant is exactly calculable in two dimensions,
it can also be done beyond perturbation theory in the Schwinger
model. This was first realized by van den Doel \cite{Doel},
and we shall here briefly review his calculation (see also 
\cite{CK,G}). At the same time we shall make contact with the very
similar calculation in four-dimensions, in
quenched chiral perturbation theory based on the replica formulation.

In the continuum theory, we consider the Lagrangian
\begin{equation}
{\cal L} = -\frac{1}{4}F_{\mu\nu}F^{\mu\nu} - \sum_{i=1}^{N_f}\bar{\psi}_i
(i\slash{\partial} + m - g\slash{A})\psi_i
- \frac{g\theta}{4\pi}\epsilon_{\mu\nu}F^{\mu\nu},
\end{equation}
with gauge group U(1), coupling $g$, and $N_f$ species of fermions.
A two-dimensional $\theta$-term has been included as well. 
Since the fermion determinant is exactly
calculable in two dimensions, the theory
is in large measure soluble.
A convenient representation
of the model is its bosonized form \cite{bosonization} where the
Lagrangian density, upon an exact integration over the gauge potential, 
takes the form
\begin{eqnarray}
{\cal L} &=& 
\sum_{j=1}^{N_f}\frac{1}{2}\partial_{\mu}\phi_j\partial^{\mu}\phi_j
-\frac{g^2}{2\pi}\left(\sum_{\j=1}^{N_f}\phi_j + \frac{\theta}{2\sqrt{\pi}}
\right)^2 \cr
&& + cm^2\sum_{\j=1}^{N_f}{\cal N}\cos(2\sqrt{\pi}\phi_j) .
\label{boso}
\end{eqnarray}
Here ${\cal N}$ denotes normal ordering and $c=e^{\gamma}/2\pi$
where $\gamma$ is Euler's constant.

Although there is no spontaneous chiral symmetry breaking here, the
bosonized action~(\ref{boso}) bears a strong resemblance to chiral
Lagrangians in four dimensions, with $f = 1/\sqrt{\pi}$ playing the
role of a (dimensionless) pion decay constant \cite {DNS}. 
This becomes particularly clear when
we consider the theory in an expansion around static (zero-momentum)
modes. If we define the partition function in a sector of fixed topological
charge $\nu$ by means of~\cite{LS}
\begin{equation}
{\cal Z}_{\nu} \equiv \frac{1}{2\pi}\int_{0}^{2\pi}\! d\theta\,
e^{i\nu\theta} {\cal Z} (\theta),
\end{equation}
the integral over $\theta$ can be performed exactly. Upon defining 
the $N_f\times N_f$ matrix $U \equiv {\rm diag}\{\exp(i2\sqrt{\pi}\phi_j)\}$,
the terms that survive in the static limit yield a Boltzmann factor
$$
     (\det U)^{\nu}\exp\left[-\frac{c}{2}Vm^2{\cal N}\,{\rm Tr}\,
(U + U^{\dagger})\right],
$$
in analogy with the result in 4 dimensions \cite{LS} (here
$V$ denotes the finite two-dimensional volume). Note,
however, that the mass term is proportional to $m^2$ rather than
to $m$ as in four dimensions. This is crucial for understanding the
difference with respect to spontaneous breaking of chiral symmetry. 

The analogy to a four-dimensional chiral Lagrangian holds to any
order in chiral perturbation theory. In 
particular it is useful for understanding the quenched limit of
the theory. The mass term and its normal-ordering prescription
complicate matters slightly, and it suffices to consider the massless
limit. In that limit we read off the diagonal scalar propagator from
Eq.~(\ref{boso}),
\begin{equation}
{\cal G}(p^2) = \frac{1}{p^2} - \frac{g^2}{\pi}
\frac{1}{p^2(p^2+N_fg^2/\pi)} \label{Nfprop},
\end{equation}
which reduces to the classic Schwinger result of an ordinary
massive propagator when $N_f=1$ \cite{CK}; in the quenched limit ($N_f=0$)
a double pole develops,
\begin{equation}
{\cal G}(p^2) = \frac{1}{p^2} - \frac{g^2}{\pi}
\frac{1}{(p^2)^2}\qquad(N_f = 0),
\end{equation}
as first noted by van den Doel \cite{Doel}.
This phenomenon is completely analogous to what happens in quenched
chiral perturbation theory in four dimensions when formulated
in terms of replicas; see the appropriate replica Feynman
rule in Eq.~(7) of Ref.~\onlinecite{DS}.
The only difference is the replacement
of the Schwinger mass parameter $\mu^2 = g^2/\pi$ 
by what is there commonly normalized
as $\mu^2/N_c$, where $N_c$ is the number of colors. (The double
pole had of course been observed much earlier in quenched
chiral perturbation theory by means of a supersymmetric extension~\cite{BG,S}.)

The above description of the diagonal scalar propagator glosses over
the fact that in two dimensions the propagation of massless degrees
of freedom require an infrared regularization. This is true even when
the Schwinger mass $\mu = g/\sqrt{\pi}$ is taken into account because of
the remaining $1/p^2$-poles in the propagator (\ref{Nfprop})
when $N_f \to 0$. After regularizing
this infrared divergence by an additional mass parameter $m_{IR}$, the
calculation of the chiral condensate requires integration over a closed
loop of the propagator
\begin{equation}
{\tilde{\cal G}}(p^2) = 
\frac{p^2+(N_f-1)\mu^2}{(p^2+m_{IR}^2)(p^2+N_f\mu^2+m_{IR}^2)} - 
\frac{1}{p^2+M^2} ,
\label{Nfpropreg}
\end{equation}
where $M$ is the arbitrary mass defining the normal-ordering prescription
\cite{bosonization}. From here
the quenched chiral condensate has been computed
\cite{Doel,CK,G} to give
\begin{eqnarray}
\langle\bar{\psi}\psi\rangle &=& 
 - \lim_{m_{IR}\to 0\atop N_f\to 0}cM\langle {\cal N} 
\cos[2\sqrt{\pi}\phi_k]\rangle \cr
&=& - \lim_{m_{IR}\to 0\atop N_f\to 0}
cm_{IR}\left(1+\frac{N_f\mu^2}{m_{IR}^2}\right)^{1/2N_f} .
\end{eqnarray}
For fixed infrared cutoff $m_{IR}$, the
limit $N_f \to 0$ yields \cite{CK}
\begin{equation}
\langle\bar{\psi}\psi\rangle = -\lim_{m_{IR}\to 0} c m_{IR} e^{\mu^2/2m_{IR}} ,
\end{equation}
which is infrared divergent. This ordering of limits seems closest to
the actual ``physical'' (i.e., computational) definition of the quenched
theory, and we shall view it as the simplest manifestation of
the difficulty with defining the quenched Schwinger model. 
Since the result is divergent, and since the computation has been
done based on an expansion around the massless theory, one can question to 
what extent the resulting divergence is a truly reliable prediction. 
More detailed
computations at fixed $V$ can be found in Refs.~\onlinecite{Smilga}
and~\onlinecite{DurrSharpe}.

\section{Monte Carlo analysis}

We consider a lattice with volume $(La)^2$ and employ a non-compact formulation for the gauge field.
In a sector with topological charge $Q$, the gauge field $A_\mu$ may be decomposed \cite{NN_U1} as
\begin{eqnarray}
A_1(x) &=& \partial^*_2 \phi(x) + \frac{2\pi}{L} h_1 - \partial_1 \alpha(x), \\
A_2(x) &=&  - \partial^*_1 \phi(x) + \frac{2\pi}{L} h_2
 - \partial_2 \alpha(x) - \frac{2\pi Q}{L^2}. 
\end{eqnarray}
Here $\phi(x)$ is a real periodic function with
no zero mode, and $h_\mu$ are two real constants in the
interval $(-1/2, 1/2]$ that parametrize the two Polyakov loops on the
2-d lattice.  
$\partial_\mu$ and $\partial^*_\mu$ are forward and backward
finite differences; $\alpha(x)$ represents the gauge
degree of freedom. The gauge action is then
\begin{eqnarray}
S_G &=& \frac{1}{4g^2} \sum_x F^2_{\mu\nu}(x) \\
&=& \frac{1}{2g^2} \sum_x \left[ \Delta \phi(x) \right]^2 +
 \frac{2\pi^2 Q^2}{(g L)^2} ,
\label{eq:action}
\end{eqnarray}
where
\begin{equation}
\Delta \phi(x) = \sum_\mu \left[ \phi(x+\mu) + \phi(x-\mu) - 2 \phi(x) \right] .
\end{equation}
We denote the gauge coupling by $\beta=1/(2g^2 a^2)$.
The continuum limit may be taken at fixed finite $g$ and at fixed volume by taking $\beta$ and~$L$ to infinity.  Alternatively, we can keep $g$ and~$L$ fixed and vary the physical volume by changing $\beta$.

Our choice of the non-compact action is motivated mainly
by the ease with which we can generate independent gauge configurations with the desired topological charge.
Successive configurations are generated by a heat bath so that there is no autocorrelation.
We restrict ourselves to the sector with zero topological charge.

We employ lattices of linear size $L$ ranging from~8 to~60, and fix $\beta=2$.
We also have one data set with $\beta=1$ and $L=48$; if we renormalize at fixed
$g$ as discussed above then this is equivalent to $L=48\sqrt2\simeq68$ at $\beta=2$, allowing us a larger physical volume at modest additional cost.

We use the massless overlap--Dirac operator \cite{OvlapD}
for the fermions,
\begin{equation}
D = \frac{1}{2} \left[ 1 + \gamma_5 \epsilon(H_W) \right] .
\label{eq:Dov}
\end{equation}
Here $H_W = \gamma_5D_W(-1)$ is the hermitian Wilson--Dirac operator with
mass parameter set to $-1$. With this normalization the eigenvalues
of the hermitian overlap--Dirac operator $H = \gamma_5 D$ lie in the
interval $[-1,1]$, but there is a wave-function renormalization $Z_\psi = 2$
with respect to the conventional normalization of the Dirac operator
\cite{EHN} that one needs to keep in mind.

For $L\le32$ we diagonalize $H_W$ exactly by a Householder transformation 
followed by $QL$ iteration \cite{NumRec}.  From this we construct $H$, 
whose eigenvalues we obtain similarly.
For the larger lattices we compute the lowest eigenvalues of $H^2$
with the Ritz variational algorithm of Kalkreuter and Simma~\cite{Ritz}.
Here the action of $\epsilon(H_W)$ on a vector is obtained using a 
modified version of the two-pass algorithm \cite{twopass}. In the first
pass, the Lanczos algorithm is used to obtain a tridiagonal matrix, $T_W$, 
that is a good approximation to $H_W$. Then $\epsilon(T_W)$ is obtained by
an exact diagonalization of $T_W$, and the second pass is used to compute
the action of $\epsilon(T_W)$ on a vector. 

\begin{figure*}[ht]
\begin{center}
\includegraphics*[width=1.7\columnwidth]{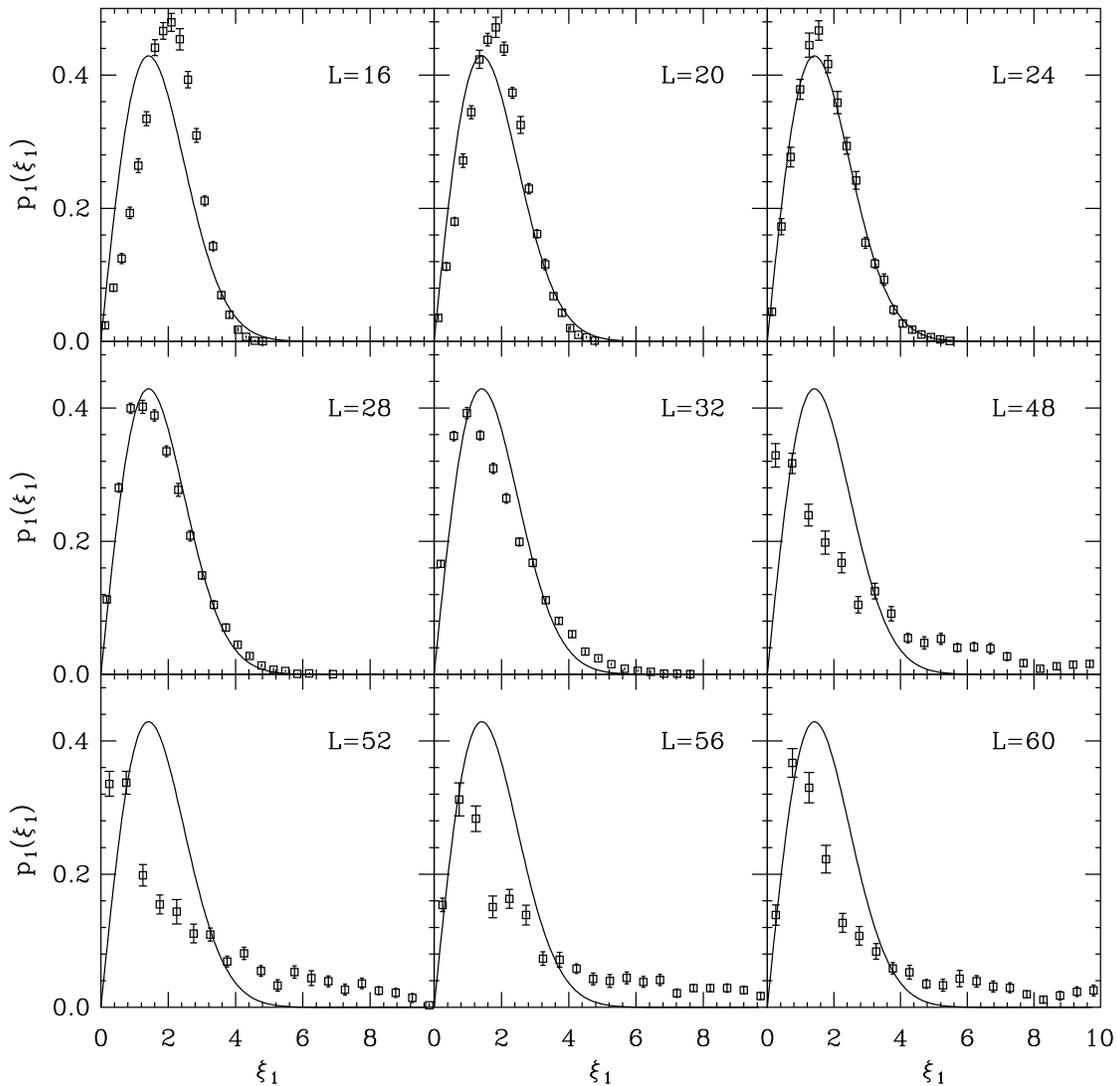}
\caption{Comparison of the distribution of the rescaled lowest eigenvalue
of the quenched Schwinger model
with the prediction from RMT for the quenched chUE. All ensembles here
have $\beta=2$.
}
\label{fig:Pmin_Nf0}
\end{center}
\end{figure*}

Previous papers have compared the eigenvalue distributions to the 
predictions of random matrix theory (RMT), and we shall proceed to do the same.
In higher dimensions, where spontaneous chiral symmetry breaking is 
allowed, it is by now well known that the lowest eigenvalues of
the Dirac operator in the $\epsilon$-regime are distributed
according to universal finite-size scaling
distributions that can be derived either from RMT
\cite{SV,ADMN} or directly from the chiral Lagrangian framework
\cite{OTV,AN}. There is ample evidence that the universality class of these 
distributions is dictated by the way
the fermions transform under the gauge group. This has been demonstrated 
for three different gauge group representations with
overlap fermions \cite{EHKN}, and for a variety of exotic 
representations with staggered fermions \cite{DHNS}. 

According to this classification, if chiral symmetry could be spontaneously
broken in the Schwinger model, where 
the fermions transform as a complex representation, the relevant
universality class in RMT terms would be that of the chiral unitary 
ensemble (chUE). It is by no means clear, however,  whether these predictions 
are of any relevance to the quenched Schwinger model, where
spontaneous chiral-symmetry breaking is prohibited \cite{MW}. 
Ref.~\onlinecite{Lang} included such a comparison, but the
results were not easy to understand. For small lattices the
eigenvalue distributions appear to fall in with the chiral symplectic 
ensemble (chSE), which
should be of relevance to {\em real\/} fermion representations.
For larger lattices there appears to be a switch to the perhaps more
natural chUE. With our greater statistics and our larger lattices, we are 
now able to see that such surprising results do not really hold.

We begin with attempts to fit the distribution
of the lowest eigenvalue to the form predicted by RMT for the quenched 
($N_f=0$) chUE \cite{DN},
\begin{equation}
p_1(\xi_1) =\frac{1}{2}\xi_1 e^{-\xi_1^2/4},
\end{equation}
where $\xi_1 = \lambda_1 \Sigma L^2$ and the condensate $\Sigma$ is the
fit parameter. We compare to histograms with 
20 bins in $\xi_1$. The comparisons are shown in Fig.~\ref{fig:Pmin_Nf0}
for nine volumes,
and the results of the fits to the RMT prediction
are listed in Table~\ref{table:Nf0_fits}.

\begin{table}[ht]
\caption{%
Number of configurations studied for each volume $V=L^2$, the condensate
$\Sigma$, the $\chi^2$, number of degrees of freedom, and confidence
level from fits of the lowest eigenvalue distribution to the RMT form
of the chUE
with 20 histogram bins. The last line is for $\beta=1$, all others
for $\beta=2$.
\label{table:Nf0_fits}}
\begin{ruledtabular}
\begin{tabular}{rrlrrr}
 $L$ & $N_{\rm meas}$ & $\Sigma$ & $\chi^2$ & dof & CL \\
\hline
  8 &  1000 & 0.2221(25) & 415.5 & 19 & $2.6 \times 10^{-76}$ \\
 12 & 10000 & 0.1705(6) & 3278. & 19 & $< 10^{-100}$ \\
 16 & 10000 & 0.1752(4) & 2801. & 19 & $< 10^{-100}$ \\
 20 & 12440 & 0.1632(5) & 1024. & 19 & $< 10^{-100}$ \\
 24 &  4880 & 0.1679(11) & 58.14 & 19 & $7.6 \times 10^{-6}$ \\
 28 &  9360 & 0.1879(11) & 204.9 & 18 & $1.1 \times 10^{-33}$ \\
 32     & 12320 & 0.2242(14) & 1223. & 19 & $< 10^{-100}$ \\
 48 &  1660 & 0.828(19) & 945.4 & 17 & $< 10^{-100}$ \\
 52 &  1280 & 1.164(16) & 714.9 & 17 & $< 10^{-100}$ \\
 56 &  1320 & 2.115(29) & 894.7 & 18 & $< 10^{-100}$ \\
 60 &  1020 & 2.36(11) & 929.8 & 19 & $< 10^{-100}$ \\
\hline
 48 &   860 & 3.921(42) & 457.4 & 17 & $1.7 \times 10^{-86}$ \\
\hline
\end{tabular}
\end{ruledtabular}
\end{table}

$L=24$ is the case that comes closest to agreement with the chUE prediction, but even here our fit
gives a $\chi^2/{\rm dof}$ of $58.1/19$ corresponding to a confidence
level of $7.6 \times 10^{-6}$. Our high-statistics data enable us to 
rule out the chUE scenario here.  As can be seen from
Table~\ref{table:Nf0_fits},
for all other lattice sizes the RMT fits to the chUE predictions are ruled out
even more thoroughly.

Farchioni et al.~\cite{Lang}
found agreement with the chUE prediction for large volumes,
in particular for $L=16$ and $\beta=1$ (see their Fig.~6; their definition of $\beta$ is twice ours).  This is the same physical volume as $L\simeq23$ and~$\beta=2$.  As we stated above, we rule out the chUE at this volume and at all other volumes studied.%
\footnote{Ref.~\onlinecite{Lang} describes a simulation of a compact gauge action, however, so the comparison between our calculations cannot be made precise.}
For smaller volumes, Farchioni et al.\ favor the chSE, claiming in particular a good fit at $L=16$ for $\beta=2$. A fit of our high-statistics $L=16$ data
to the quenched chSE distribution gives
$\chi^2/{\rm dof} = 436/19$, ruling it out as well.

As seen in Fig.~\ref{fig:Pmin_Nf0},
the peak of the distribution of the lowest eigenvalue
moves downward faster than $1/L^2$, the RMT prediction, while a sizable tail 
of the distribution persists.
The RMT scaling law is based on the assumption of a finite eigenvalue 
density $\rho(0)$ at the origin;
a scaling faster than $1/L^2$ thus entails a divergence in
$\rho(0)$ as $L\to\infty$. By the Banks--Casher relation $\Sigma = 
\pi\rho(0)$, this in turn implies a divergent chiral condensate.
Indeed, from Table~\ref{table:Nf0_fits} we see that the fitted $\Sigma$
grows quite rapidly with increasing $L$.

We now elaborate on this last point. 
Let $\lambda_i(L,\beta)$ be the
$i$-th lowest non-zero eigenvalue of $H$ on an $L^2$ lattice at a 
coupling $\beta$. For a finite chiral condensate to form we expect that
\begin{equation}
f_i(L,\beta) \equiv \frac{1}{L^2\langle \lambda_i(L,\beta)\rangle} 
\end{equation}
approaches an $i$-dependent constant as $L \to \infty$,
\begin{equation}
\lim_{L\to \infty}f_i(L,\beta) ~=~ f_i(\beta) ~.
\end{equation}
In lattice units we expect $f_i(\beta) \propto \Sigma$ \cite{LS}.
Furthermore, the products $f_i(\beta)\sqrt{\beta}$ should
approach finite continuum limits for $\beta \to \infty$. 

\begin{figure}[ht]
\begin{center}
\includegraphics*[width=.8\columnwidth]{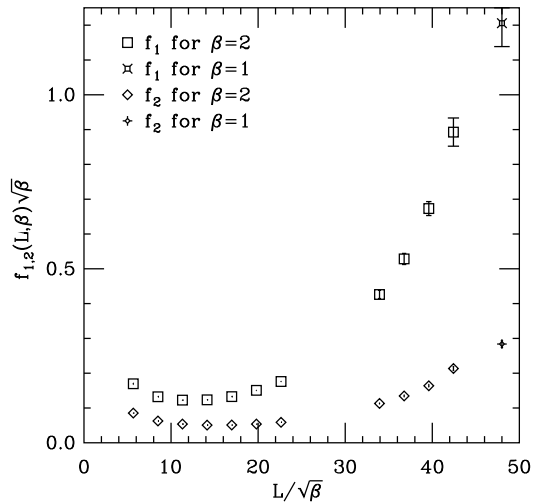}
\caption{Evidence for a diverging condensate in the quenched Schwinger model.}
\label{fig:f12}
\end{center}
\end{figure}

The scaled variables $f_{1}(L,\beta) \sqrt{\beta}$ and 
$f_{2}(L,\beta) \sqrt{\beta}$ are plotted as functions of
the physical size $L/\sqrt{\beta}$ in 
Fig.~\ref{fig:f12}.  (We combine data for $\beta=1.$ with data for 
$\beta=2.$)
We see that these quantities do not approach $L$-independent constants
as one would expect on
the basis of the existence of a finite chiral condensate $\Sigma$.

\begin{figure*}[ht]
\begin{center}
\includegraphics*[width=1.7\columnwidth]{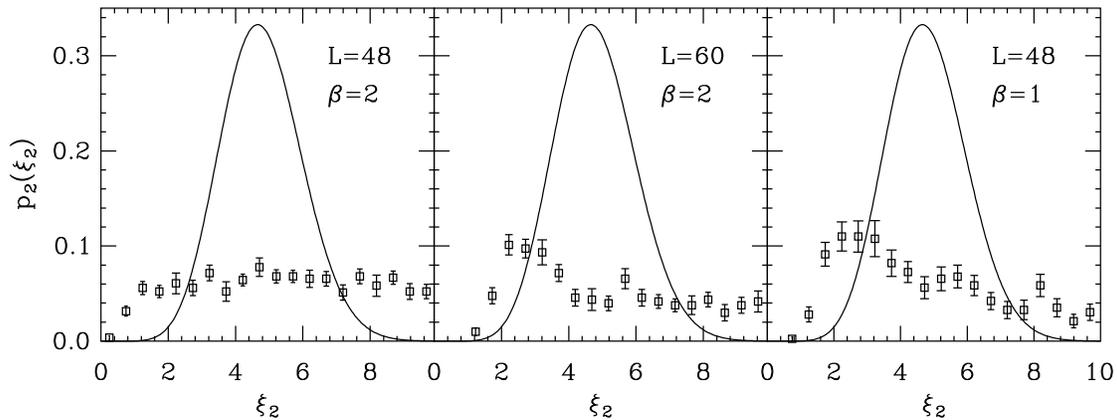}
\caption{
Distribution of the scaled second eigenvalue of the quenched Schwinger
model.
\label{fig:p2}}
\end{center}
\end{figure*}

\begin{figure*}[ht]
\begin{center}
\includegraphics*[width=1.7\columnwidth]{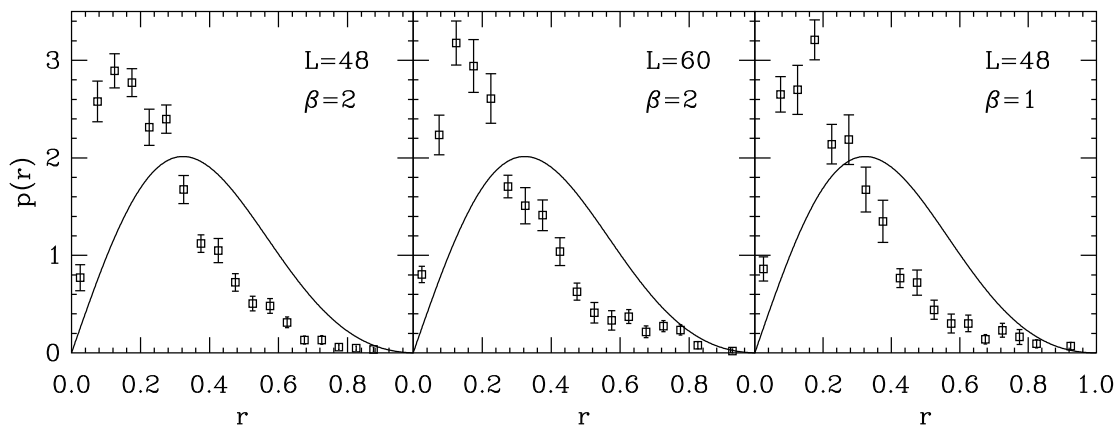}
\caption{
The distribution of
the ratio of the two lowest eigenvalues in the quenched Schwinger model
shows that they do not obey the
universal distribution given by the chUE.
\label{fig:pr}}
\end{center}
\end{figure*}

The distribution $p_2(\xi_2)$
of the {\em second\/} scaled eigenvalue (with $\Sigma$ taken from
Table~\ref{table:Nf0_fits})
for some large volumes are shown
in Fig.~\ref{fig:p2} and compared with the predictions for the ($N_f=0$)
chUE \cite{DN},
\begin{equation}
p_2(\xi_2) = \frac{1}{4} \xi_2 e^{-\xi_2^2/4} \int_0^{\xi_2} du\, u
 \left[ I^2_2(u) - I_1(u) I_3(u) \right] .
\end{equation}
Much like the distributions of $\xi_1$ above, these
clearly do not fall in the
universality class of the chUE\@. We can eliminate the scale
$\Sigma(L,\beta)$ in these comparisons, by plotting
the distribution of $r={\lambda_1(\beta,L)}/{\lambda_2(\beta,L)}$.
We compare this to the prediction from the chUE \cite{DN,NNN}
\begin{eqnarray}
p(r) &=& \frac{1}{4} \frac{r}{(1-r^2)^2} \int_0^\infty du \,u^3
 \exp\left(-\frac{u^2}{4(1-r^2)}\right)\nonumber\\
 &&\quad\times \left[ I^2_2(u) - I_1(u) I_3(u) \right] ,
\end{eqnarray}
in Fig.~\ref{fig:pr}.
Again, the data do not fall in the universality class of the chUE.


\begin{figure}[ht]
\begin{center}
\includegraphics*[width=.8\columnwidth]{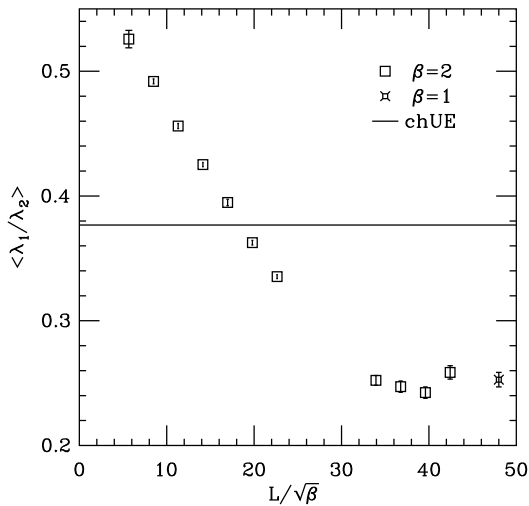}
\caption{
Expectation value of $r={\lambda_1(L,\beta)}/{\lambda_2(L,\beta)}$
in the quenched Schwinger model
as a function of the physical size. The data are not
consistent with chUE but do seem to approach a finite limit as $L\to\infty$.
}
\label{fig:ravg}
\end{center}
\end{figure}

Do the $p_i(\xi_i)$ reach some limiting distributions as $L\to\infty$? It is 
difficult to answer this question based on the data shown in
Figs.~\ref{fig:Pmin_Nf0} and \ref{fig:p2}. It is conceivable that 
$p_1(\xi_1)$ approaches a function peaked at zero while 
$p_2(\xi_2)$ reaches a limiting form peaked away from zero, 
but we have no real evidence for this.
Whatever the answer, however, it is possible that the distribution 
$p(r)$ of the ratio {\em does\/} have a limiting distribution. 
Even though $\lambda_1$ and $\lambda_2$ go to zero much
faster than $1/L^2$, we can discern some level repulsion that
favors a ratio $r$ between~0 and~1.
Figure~\ref{fig:ravg} shows $\langle r \rangle$ as a function of
the physical size and this average seems to approach a 
finite limit as $L\to\infty$.


\section{The unquenched theory}

When we compute all the eigenvalues of the Dirac operator on each pure gauge
configuration, we can calculate observables in
the theory with $N_f \neq 0$ by reweighting with the fermion 
determinant. We restrict
ourselves to $N_f=1$ since statistical fluctuations are worse when the 
target ensemble is farther from the original quenched ensemble.
In the continuum $N_f=1$ is of course Schwinger's original model 
\cite{Schwinger}, which is exactly soluble. In particular, 
the infinite-volume chiral condensate can be computed analytically,
\cite{bosonization}
\begin{eqnarray}
\Sigma &=& \frac{e^{\gamma}\mu}{2\pi} =
\frac{ge^{\gamma}}{2\pi^{3/2}} \nonumber\\
&=& (0.1599\ldots)\,g,
\label{eq:cond}
\end{eqnarray}
in the conventional normalization,
where $\mu = g/\sqrt{\pi}$ is the Schwinger mass and 
$\gamma$ is Euler's constant.
Because the chiral symmetry is broken explicitly,
the analysis of Leutwyler and Smilga \cite{LS} and the whole RMT analysis
for the $N_f=1$
theory apply directly here. We therefore know the complete
microscopic spectrum of the Dirac operator in the $\epsilon$-regime and
it belongs to the universality class of the chUE.

\begin{table}[ht]
\caption{%
Number of configurations studied for each volume $V=L^2$ with $\beta=2$,
reweighted to $N_f=1$, the condensate $\Sigma$, the $\chi^2$, number of
degrees of freedom and confidence level from fits of the lowest eigenvalue
distribution to the RMT form with 20 histogram bins.
\label{table:Nf1_fits}}
\begin{ruledtabular}
\begin{tabular}{rrlrrr}
 $L$ & $N_{meas}$ & $\Sigma$ & $\chi^2$ & dof & CL \\
\hline
 12 & 10000 & 0.2168(6) & 2158. & 19 & $< 10^{-100}$ \\
 16 & 10000 & 0.2019(5) & 2067. & 19 & $< 10^{-100}$ \\
 20 & 12440 & 0.1781(5) & 791.8 & 19 & $< 10^{-100}$ \\
 24 &  4880 & 0.1694(12) & 60.48 & 19 & $3.2 \times 10^{-6}$ \\
 28 &  9360 & 0.1671(11) & 45.98 & 18 & $3.0 \times 10^{-4}$ \\
 32 & 12320 & 0.1648(11) & 16.73 & 19 & 0.61 \\
\hline
\end{tabular}
\end{ruledtabular}
\end{table}

Once again, the simplest quantity with which to compare is the distribution
of the smallest (non-zero) Dirac eigenvalue \cite{DN},
\begin{equation}
p_1(\xi_1) = \frac{1}{2}\xi_1 I_2(\xi_1)e^{-\xi_1^2/4},
\end{equation}
here restricted to the massless case.
\begin{figure}[ht]
\begin{center}
\includegraphics*[width=.9\columnwidth]{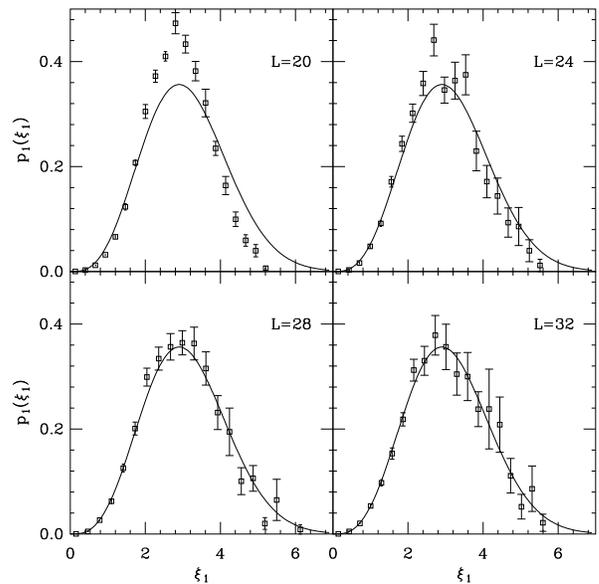}
\caption{Comparison of the distribution of the rescaled lowest eigenvalue
for $N_f=1$ with the prediction of RMT for the $N_f=1$ chUE.
}
\label{fig:Pmin_Nf1}
\end{center}
\end{figure}
\begin{figure}[ht]
\begin{center}
\includegraphics*[width=.9\columnwidth]{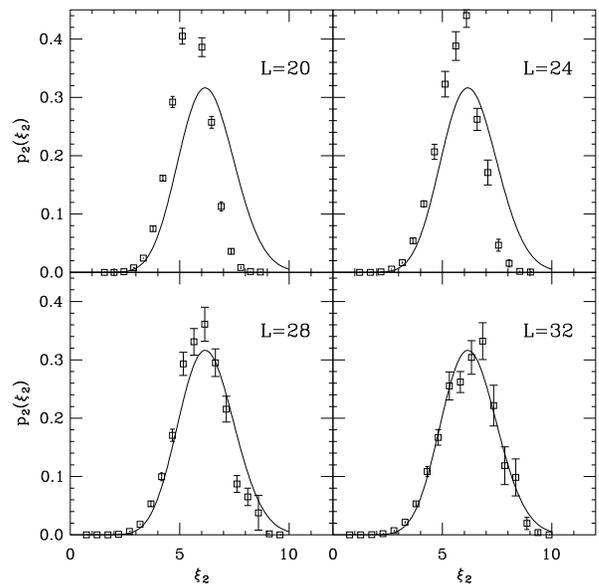}
\caption{Rescaled second eigenvalue
for $N_f=1$, compared with the prediction of RMT for the $N_f=1$ chUE
[see Eq.~(15) in Ref.~\onlinecite{NNN}].
}
\label{fig:P2_Nf1}
\end{center}
\end{figure}
We were able to reweight ensembles on lattices up to size $L=32$. Again,
we made fits to the RMT prediction, with $\Sigma$ the fit parameter,
using histograms with 20 bins. The fits are detailed in
Table~\ref{table:Nf1_fits} and shown in Fig.~\ref{fig:Pmin_Nf1}.
The agreement with the chUE is quite good for
the two largest lattices.
The second eigenvalue (Fig.~\ref{fig:P2_Nf1}) tells a similar story.
We show the averaged
ratio $\langle r\rangle$ in Fig.~\ref{fig:ravg1}, which may be
compared to Fig.~\ref{fig:ravg} for the quenched theory.
The result for $L=32$ agrees with the prediction of the chUE, $\langle r\rangle
=0.5044$ \cite{NNN}.
\begin{figure}[ht]
\begin{center}
\includegraphics*[width=.8\columnwidth]{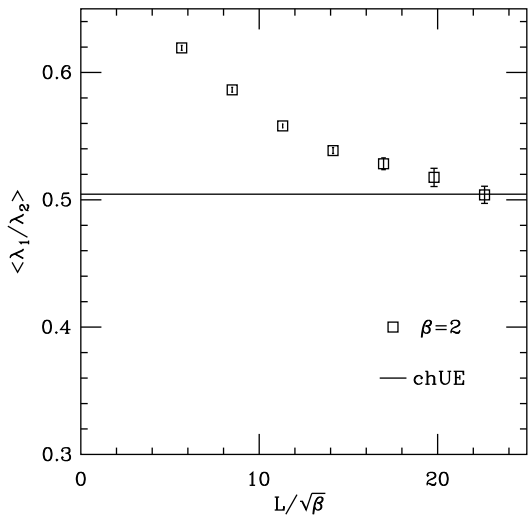}
\caption{
Expectation value of $r={\lambda_1(L,\beta)}/{\lambda_2(L,\beta)}$
for $N_f=1$
as a function of the physical size.
}
\label{fig:ravg1}
\end{center}
\end{figure}

Converting the result from Table~\ref{table:Nf1_fits} for the largest
lattice size to the conventional normalization, we find
$\Sigma/g = 0.1648(11)$, within 3\% of 
the continuum value given in
Eq.~(\ref{eq:cond}); finite lattice spacing corrections, expected
to be of ${\cal O}(1/\beta)$, could easily explain the difference (see also~\cite{NNV}).%
\footnote{The fit to $p_2(\xi_2)$ for $L=32$
gives  a value of $\Sigma$ that is 7\% larger than the value
shown in Table~\ref{table:Nf1_fits}.  This is consistent with wide
experience that higher eigenvalues suffer larger finite-volume effects.}
We note, however, that fits to the RMT distribution
of the lowest eigenvalue do not work as well for the smaller 
lattices, $L \leq 24$, giving unacceptably small confidence levels.
It would be interesting to see whether the
agreement with RMT persists on larger lattices and whether the continuum
value for $\Sigma$ is correctly reproduced as $\beta \to \infty$. 
Unfortunately, reweighting becomes prohibitive for larger volumes.
A direct numerical simulation
of the $N_f=1$ theory is probably needed.

\section{Summary}

As we have shown, the quenched Schwinger model does not fall into
any of the three universality classes of chiral symmetry breaking.
In view of the Coleman-Mermin-Wagner theorem, which forbids spontaneous
breaking of a continuous symmetry in two dimensions, this in itself
is not very surprising. The only possible loophole out of this
argument would be to note that the quenched theory is non-unitary and
thus it might not satisfy some assumptions of the theorem.

The application of the Coleman-Mermin-Wagner 
theorem to the quenched theory may be 
considered in two ways. In the
so-called supersymmetric formulation of quenching, spontaneous
chiral symmetry breaking is associated with quenched Goldstone
bosons and Goldstone fermions. In the replica formulation, the
spontaneous breaking is associated with Goldstone bosons alone.
In both ways of considering quenching, the theorem
seems to exclude rigorously the possibility of a non-zero condensate. 
In that light it
is perhaps surprising that the numerical evidence now points towards an 
ill-defined, divergent chiral
condensate, rather than a vanishing condensate.

The Schwinger model with $N_f=1$ is on an entirely different
footing due to the explicit breaking of chiral symmetry
by the anomaly. Here we have unambiguous analytical predictions
for the behavior of Dirac operator spectra near the origin, and
our $N_f=0$ results re-weighted with the Dirac determinant to
simulate the $N_f=1$ theory are consistent with these 
analytical predictions. For the {\em massive\/} Schwinger model with $N_f=1$, 
a disagreement with RMT must appear as the mass is taken to infinity.
Our statistics have not been good enough to
attempt the even more ambitious reweighting to simulate the
$N_f=2$ theory. Also here unusual results should appear, presumably
with the Dirac operator eigenvalues being strongly repelled by
the origin so as to produce a vanishing $\rho(0)$. For $N_f=2$ analytical 
calculations \cite{Smilga} suggest a behavior 
\begin{equation}
\rho(\lambda) ~\sim~ \lambda^{1/3} ~, \label{Nf2}
\end{equation}
and an interesting question is whether such behavior at the
rescaled level has a universal distribution from ``critical''
Random Matrix Theories (the precise behavior (\ref{Nf2}) is actually
realized in a very simple chiral matrix model
\cite{Janik}) or analogous eigenvalue models
\cite{Gernot}. This could
be an interesting topic for future investigations.

\vspace{1cm}
PHD and UMH would like to thank the Kavli Institute of Theoretical
Physics for its hospitality and NSF grant PHY99-07949 for partial
support during the completion of this work. 
The work of BS was supported in part by the Israel Science Foundation 
under grant no.~222/02-1.
The work of PHD, UMH, and BS was also supported by NATO Collaborative 
Linkage Grant PST.CLG.977702.
The bulk of our calculations were carried out on computers provided by the 
High Performance Computing Unit of the Israel Inter-University Computation 
Center.
Additional computer resources were provided by
Jefferson Lab 
and by the Tel Aviv University School of Computer Science.
Thanks also to the NorduGrid collaboration for providing support
and middleware.
RN thanks the Niels Bohr Institute for its hospitality and
acknowledges partial support by the NSF under grant number PHY-0300065,
and also partial support from Jefferson Lab, operated by SURA under
DOE contract DE-AC05-84ER40150.

\end{document}